%% file: ms.tex
\input epsf.tex
\documentclass[preprint2]{aastex}
\shorttitle{Binaries in $\rho$ Ophiuchus}
\shortauthors{Barsony, Koresko, \& Matthews}
\begin{document}
\title{\rm A Search for Close Binaries in the $\rho$ Ophiuchus Star-Forming Region}
%
\author{M. Barsony\altaffilmark{1,2}}
\affil{Jet Propulsion Laboratory, Mail Stop 169-327 \\ 
4800 Oak Grove Drive, Pasadena, CA  91109}
\email{fun@uhuru.jpl.nasa.gov}

\author{C. Koresko\altaffilmark{2}}
\affil{Michelson Science Center\\
California Institute of Technology\\
1201 E. California Blvd., Pasadena CA 91125}
\email{koresko@fitzgerald.jpl.nasa.gov}

\and

\author{K. Matthews}
\affil{Caltech Optical Observatory\\
California Institute of Technology \\
Pasadena, CA 91125}
\email{kym@caltech.edu}

\altaffiltext{1}{and Space Science Institute, 
3100 Marine Street, Suite A353, Boulder, CO  80303-1058}
\altaffiltext{2}{Based on observations obtained at the Hale telescope, 
Palomar Observatory, as part of a continuing collaboration between
the California Institute of Technology, NASA/JPL, and Cornell
University.}



\begin{abstract}

We have carried out a new, near-infrared speckle imaging survey of 19
members of the young stellar population in the nearby (d$=$140 pc), $\rho$
Ophiuchi cloud core. Results for four binary and one newly discovered
triple system are reported. Data for all known multiple  systems among the
pre-main-sequence population of $\rho$ Oph are tabulated. We define a {\it
restricted binary fraction} $F_{b,r}$ and a {\it restricted companion
fraction} $F_{c,r}$ as counting only those systems most detectable in the
present and previous high-resolution near-infrared imaging surveys, having
separations between 0.\arcsec1 and $\sim$ 1.\arcsec1 and K-band magnitude
differences  $\Delta K < 3$.  Analysis of all the available multiplicity
data results in updated values of $F_{b,r}$ = 24\% $\pm$ 11\% and
$F_{c,r}$ = 24\% $\pm$ 11\% for the Ophiuchus pre-main-sequence
population.  These values are consistent with the values in the Taurus
star-forming region, and $F_{c,r}$  is in excess by a factor of 2 relative
to the Main Sequence at the 1 $\sigma$ level.

\end{abstract}




\section{Introduction}

The formation of binary and multiple stars is a natural consequence of
present-day star formation, yet many basic questions regarding this
process remain unanswered.  What is the frequency of multiple  systems in
the nearest star-forming regions to Earth?   What is the distribution of
source separations among pre-main-sequence binaries? Does this
distribution evolve with time as the stars join the Main Sequence field
population? Does the binary separation distribution vary with disk
parameters, and if so, how? Do close binaries affect the appearance of
each individual disk, and if so, how ? 

It has been suggested that the binary frequency among T-Tauri stars in
Taurus and Ophiuchus  is higher than among the nearby solar-like stars in
a restricted separation range (Ghez, Neugebauer, \& Matthews 1993; Leinert
et al. 1993).   This separation range is 0\arcsec.1-1\arcsec.8, 
corresponding to projected linear separations of 14-250 AU at an assumed 
140 pc distance to the Taurus and Ophiuchus star-formation regions.
Subsequent investigations of this claim have found conflicting, or at
least, ambiguous, results for different star-forming regions (e.g.,
Prosser et al. 1994; Simon et al. 1995; Beck, Simon, \& Close 2003).  In
Ophiuchus, in particular, there have been too  few known multiple systems,
with too few systems searched with comparable techniques, to have enough
statistics for comparison with the main-sequence population and with other
star-forming regions.

One problem, until recently, had simply been that too few actual cloud
members were known in $\rho$ Oph (e.g., Wilking, Lada, \& Young 1989).
This is due to the high overall extinction towards the $\rho$ Oph cloud
core, in contrast to the much lower extinction towards members of the
young stellar population in Taurus. The advent of recent satellite data
identifying young stellar objects (YSOs) in $\rho$ Oph from the X-ray
(Casanova et al. 1995; Grosso et al. 2000; Imanishi, Koyama, \& Tsuboi
2002) and the mid-infrared (Bontemps et al. 2001) wavelength regions has
significantly increased the known number of cloud members.  Large-area,
magnitude-limited, near-infrared surveys of $\rho$ Oph  have recently
become available (Barsony et al. 1997; 2MASS\footnote{2 Micron All Sky
Survey Second Incremental Release Point Source Catalog, available at {\tt
http://irsa.ipac.caltech.edu}}).  The vast majority ($\sim$ 95\%) of the
near-infrared sources are  background objects.  However,  positional
cross-correlation of the near-infrared sources with lists of known cloud
members, arrived at through observations at X-ray, mid-infrared, or
optical wavelengths, allows the identification of suitable targets that
meet the joint criteria of cloud membership and being bright enough at
K-band (2.2 $\mu$m) for speckle imaging.

Once cloud members have been identified, a survey for multiplicity should
ideally be undertaken such that all targets are searched using uniform
criteria. The lack of consistency of previous surveys has made it
difficult to draw firm conclusions regarding the binary fraction in $\rho$
Oph. Previous multiplicity surveys in $\rho$ Oph have used a variety of
search techniques, including spectroscopy and imaging, at different
wavelengths (either  near-infrared or optical), with different
sensitivities,  valid over different separation ranges. One would ideally
use a single technique, with the same sensitivity for all targets
searched, obviating the need for various corrections due to lack of
uniform data (e.g. Duchene 1999).

The largest number of binary identifications among pre-main-sequence
members of the Taurus-Auriga dark clouds were found by high-resolution
near-infrared imaging techniques (Ghez, Neugebauer, \& Matthews 1993;
Leinert et al. 1993; Simon et al. 1995). In practice, these techniques
require relatively bright targets (K $\le$ 8.5 on a 5-meter aperture
telescope for speckle imaging).  

We have undertaken a new near-infrared speckle imaging survey of {\it bona
fide} $\rho$ Oph cloud members at the Hale 5-meter telescope. Since the
{\it Chandra}, {\it ROSAT},  and {\it ISOCAM} surveys covered rather
limited areas, of order 35$^{\prime}$ $\times$ 35$^{\prime}$, containing rather
few bright K-band targets that had not previously been searched, we
supplemented our target list with bright K-band sources that are optically
identified cloud members, either from large-area ground-based H$\alpha$
surveys (Struve \& Rudjkobing 1949; Dolidze \& Arakelyan 1959;  Wilking,
Schwartz, \& Blackwell 1987) or via optical follow-up imaging of  {\it
Einstein}-identified cloud members \cite{boa92}.

We report our target list, observing procedures, and data reduction
methods in $\S$2. The results of this speckle survey are presented in
$\S$3. In the Discussion ($\S$4), we have collected published data on 49
separate multiple systems in Table 3,  and the data on all systems found
to be singles via high-resolution near-infrared techniques in Table 4, in
order to be able to draw valid, statistically significant conclusions
about the binary separation distribution and the binary fraction in $\rho$
Oph. We summarize our findings in $\S$5.

\section{Observations and Data Reduction}

Speckle observations of the 19 pre-main sequence stars in the sample were
made in the near-infrared K band on 24-25 May 2002 at the 5~m Hale
telescope on Palomar Mountain.  An external optical system (re-imager)
magnified the telescope platescale to produce a pixel size of 0\arcsec.034
$\pm$ 0\arcsec.001, sufficient to Nyquist sample the 0\arcsec.1
diffraction limit of the telescope  (Weinberger 1998).  The detector was
D-78, a camera at the Cassegrain focus which was equippped with a
$256\times 256$ pixel InSb array, of which we used the central 64 $\times$
64 pixels, giving a field-of-view of 2\arcsec.2.  The orientation accuracy
of the detector is 0.5$^{\circ}$. The seeing was typically 0\arcsec.75
FWHM. The presence of thin  cirrus prevented us from obtaining absolute
photometry.

The observing procedure was typical for speckle interferometry,  with the
exception that the chopping secondary was used to switch  rapidly between
the position of the science target and the blank  sky 30\arcsec\ away. 
Briefly, the data consisted of series of 560  frames with 0.1 second
exposure times--fast enough to ``freeze''  the atmospheric seeing and
preserve diffraction-limited image information.  
Sky exposures were interleaved with the exposures on the science
target.  In total, each series contained  490 frames on the target and 60
on the sky position, with the remaining 10 being rejected 
while the secondary's position was changing. At least 8
such series  were obtained on each science target, along with a similar 
number on an unresolved reference star. The reference-star  series were
interleaved with the target-star series to  reduce sensitivity to seeing
and focus changes.  A log of the observations is given in Table 1.

Reduction of the data consisted of the usual flatfielding  and
sky-subtraction of the raw frames, followed by computation of the Fourier
power spectrum and bispectrum (Lohman, Wiegelt, \& Wirnitzer 1983) 
averaged for the on-source frames in each series.  The power spectrum of
the calibrated sky frames in the series  was subtracted from that of the
on-source frames, and the Fourier phase  for the series was reconstructed
from the bispectrum using the recursive technique. The results for the
science target series were calibrated pairwise  with the results for the
reference-star series to remove the  distortions due to the atmosphere and
the telescope.   The calibration consisted of simple division of the
sky-subtracted  Fourier power spectra and subtraction of the reconstructed
Fourier phases.   The calibrated Fourier transform data were then averaged
over  the series for each science target.
Finally, the calibrated  Fourier data were apodized to
approximately diffraction-limited  resolution and Fourier transformed to
recover a calibrated high-resolution image.

Basic parameters (the separation, position angle, and brightness relative
to the primary) for the companions in the detected binary and triple
systems were derived by fitting models to the calibrated Fourier data. 
Parameter uncertainties were estimated by fitting simulated data
representing binaries with parameters close to the observed values.  The
simulated data were created by calibrating reference star files against
each other to produce Fourier power spectra and phases for point sources. 
Their noise should be very similar to that for the calibrated science
target data.  The modulations corresponding to double-stars were then
impressed onto the point-source power spectra by multiplying them by the
power spectra of model doubles, and onto the point-source phases by adding
the phases of the model doubles.  In total, 23 independent simulated
Fourier transforms were constructed for each real double detected in our
survey.  The results of fitting these simulated data could then be
compared directly to the true model parameter values.  The values we quote
are the root-mean-squared scatter among the fits to the simulated
datasets.  We found that the separation error was independent of the
separation and close to 0\arcsec.03 for all doubles, while the position
angle error ranged from $\sim 0.1^{\circ}$ for separation~0\arcsec.7 to 
20$^{\circ}$
for the marginally-resolved double with separation 0\arcsec.04.  The
brightness ratio errors (with the ratio defined as the brightness of the
fainter star divided by that of the brighter one) are $\sim 0.03$ for the
fully-resolved doubles and 0.15 for the marginally-resolved double. These
error estimates do not include the uncertainties in the image scale and
orientation of the detector.  Further, there is a 180$^{\circ}$ ambiguity
in the position angle of the marginally-resolved double in the ROXs 47A
hierarchical triple system. 

\section{Results}

Of the 19 target objects for which we acquired near-infrared speckle data,
five were resolved into multiple systems. Figures 1-5 show images related
to each multiple system, with three frames plotted for each of the 
five panels.
The leftmost panels represent the calibrated power spectra and the middle
panels represent the Fourier phases, both
plotted over spatial scales ranging from ~1.1\arcsec at the circles'
centers, corresponding to
approximately half the field-of-view of the portion of the detector that
was used, to the diffraction limit of ~0\arcsec.1, at the circles' outer
peripheries. The rightmost frame in
each panel shows the final reconstructed image for each target 
object.  Only one system, the ROXs 47A triple system, is a newly reported multiple (see Table 2).

In Table 2 we present the parameters of the systems shown in Figures 1-5.
Column 1 lists the  target's name, column 2, the component separations in
arcseconds, column 3, the position angle, E of N, relative to the brighter
object at K, and the last column lists the flux ratio between the
components at K.  For reference, we also tabulate previously reported
binary parameters, when available, in Table 2.

Comparison with the previously published data for four systems in Table 2
shows no discernible change in separations and position angles for the
ROXs 2, IRS 2, and ROXs 29 binary systems. There is quite a large change
in the ROXs 42C hierarchical triple system, however. Between 8 July 1990
and 24 May 2002, the projected source separation increased by 0\arcsec.12
(corresponding to 17 AU at a distance of 140 pc), whilst the position
angle has advanced by 18$^{\circ}$. 


\section{ Discussion }

\subsection{Distribution of Binary Separations}

The distribution of binary separations is of interest to identify any
possible evolution of this quantity with age when compared with Main
Sequence stars, as well as to compare this distribution among different
star-forming environments. The original motivation for examining the
pre-main-sequence distribution of binary separations was to test the
hypothesis that the apparent excess of pre-main-sequence multiple systems
observed in Taurus and Sco-Oph, reflects a distribution in the
pre-main-sequence phase which is more strongly peaked in the range of
separations to which the surveys are most sensitive ($\sim 0\arcsec.1 \le
a \le 1\arcsec.8$), rather than a true excess of multiples.  In this
scenario, there would be an evolution of the shape of the the distribution
of binary separations with time, with the distribution becoming flatter by
the time the stars reach the Main Sequence (Ghez et al. 1993).

In order to have a statistically significant sample for the  determination
of the distribution of binary separations among the young stellar
population in $\rho$ Oph, we have compiled a list of all the multiple
sources so far identified in the literature, in addition to the multiple
systems identified in the study reported here. This compilation is
presented in Table 3.  The first column of Table 3 lists each young
stellar object's name, with some common alternate names for the object
listed in the last column. Note that the designation ROX refers to ``Rho
Oph X-ray'' source as detected by the {\it Einstein} Observatory
(Montmerle {\it et al.} 1983), whereas ROXs refers to an optically
detected object that falls within the rather large  {\it Einstein} X-ray
source error circle. This distinction is important, because in some cases
more than one optical counterpart is associated with a single {\it
Einstein} source (Bouvier \& Appenzeller 1992).

Since, unfortunately, it is not unusual for a single source to have a 
dozen or more names in $\rho$ Oph, for ease of identification and for
reference, coordinates are necessary. We have therefore included J2000
coordinates, good to  $\pm$ 0\arcsec.2, from the 2MASS database (unless
otherwise  indicated) for each object in the second and third columns of
Table 3. The fourth column of Table 3 indicates the type of multiplicity:
whether  the system is binary, spectroscopic binary, triple, or quadruple.
Columns 5 and 6 list the component separations in arcseconds, and the
position angle between the components, measured East of North, from the
brighter source at K, respectively. Column 7 lists the reference  from
which the multiplicity data were gleaned, with the first listed reference
being the one from which the data in columns 5 \& 6 are taken. Finally, in
addition to alternate common names for the given target, comments are also
included in the last column of Table 3.

Among the 49 independent multiple systems listed in Table 3, there are 62
separations, which are plotted in the histogram of Figure 6. This
represents an eight-fold increase in the number of  separations available
for study of the pre-main-sequence population in $\rho$ Oph since the last
such published histogram (e.g., in Simon et al. 1995). The data in Figure
6 are binned in intervals of log~$P$, as in previous authors' works
(Duquennoy \& Mayor 1991; Simon et al. 1995). The overplotted curve is the
separation distribution based on the period distribution for main-sequence
Solar-type field stars  (Duquennoy \& Mayor 1991). The conversion from a
period to  a separation distribution uses the same assumptions as previous
authors, i.e., two 0.5 M$_{\odot}$ stars in a circular orbit at 140 pc
distance (Simon et al. 1995). The vertical scale of the Main Sequence
distribution is chosen to correspond to the same integrated number of
sources as went into the histogram (62).

We note that this sample is reasonably complete over the  $\sim 0\arcsec.1
\le a \le \sim 2\arcsec$ projected separation interval, but may be
seriously incomplete at the closest  separations (where spectroscopic
searches are necessary) and at the larger ($\ge$ 10\arcsec) separations,
where few systematic searches have been undertaken, and  where
contamination by background objects becomes an issue.

There are not yet enough individual components of the multiple systems
listed in Table 3 that have classifications available as to whether they
are WTTS (weak-lined T-Tauri stars), lacking accretion disks, or CTTS
(classical T-Tauri stars), with accretion disks, to draw meaningful
conclusions from the comparison of the WTTS vs. CTTS binary separation
distributions. Such a comparison would be interesting for the study of the
effects of binarity on disk evolution.  In this context, we note that a
recent {\it HST} spectroscopic study of a sample of 20 close ($\le$
1\arcsec) binaries in Taurus  found that 4 (20\%) turn out to be in mixed
CTTS/WTTS pairs \cite{ha03}. Furthermore, there is a strong selection
effect against finding spectroscopic binaries among the CTTS, i.e.,
against finding CTTS at the closest separations, since the veiling and
emission lines characteristic of CTTS can easily overwhelm the
photospheric absorption lines that are generally used to find
spectroscopic binaries. With these caveats, however, hopefully, the
individual components of the multiple systems in Table 3 will be
characterized in the near future, and the samples will become large enough
to allow just such a comparison.

\subsection{Binarity and Multiplicity Fraction of the $\rho$ Oph 
Pre-Main-Sequence Population}

An accurate determination of the multiplicity fraction of the  young
embedded population in $\rho$ Ophiuchus presumes, first of all, that our
search list is restricted to {\it bona fide} association members
(excluding foreground or background objects).  In practice, this  means
that each target object have one or more indicator of YSO  (young stellar
object) status, such as bright X-ray emission,  broad H$\alpha$ equivalent
width,  photospheric Li absorption, associated nebulosity, high percentage
optical polarization, infrared excess, and/or  a distance determination
placing it at the cloud's distance of $\sim$ 140 $\pm$ 20 pc.

A second requirement for an accurate determination of the multiplicity is
knowledge of the number of systems that have been searched for
multiplicity by various authors, but found to be single.  We want the
search sensitivities to be as uniform as possible. Therefore, we restrict
this analysis to objects searched for multiplicity with near-infrared
techniques that are generally sensitive to source separations
0\arcsec.1$\le r \le \sim$1.1\arcsec, and  $\Delta K \le$3 mag. These joint
criteria have the added benefit of effectively excluding background
objects  that may be chance projections towards the same line-of-sight as
the target objects.

In Table 4, therefore, we list all of the target objects searched for
multiplicity by searches which would have been sensitive to detecting
binaries  with source separations in the range 0\arcsec.1$\le r \le
\sim$1.1\arcsec, and  with component magnitude differences, $\Delta K \le$3
mag. These surveys used various techniques, such as speckle observations, 
as reported here and elsewhere (Ghez et al. 1993, Ageorges et al. 1997),
lunar occultation measurements (Simon et al. 1995), and shift-and-add
imaging data (Costa et al. 2000, Haisch et al. 2002).   In Table 4, each
target object's name is listed in the first column, its J2000 coordinates,
from the 2MASS Second Incremental Release Point Source Catalog, are listed
in the second and third columns, and the authors who have searched each
target for multiplicity and found it to be single, are listed in the last
column. In constructing the Table, we were careful not to count the same
target  object multiple times, even when referred to by different names
by  the different surveys.

Comparison of the data presented in Tables 3 \& 4 for the $\rho$ Oph cloud
core with previous work on $\rho$ Oph, other star-forming regions, and the
Main Sequence, requires determination of the {\it binary fraction},
$F_b={{B+T+Q} \over {S+B+T+Q}}$, and the {\it companion star fraction},
$F_c={{B+2T+3Q} \over {S+B+T+Q}}$, where $S$ is the number of  single
stars, $B$ is the number of binary systems, $T$, the number of triple
systems, and $Q$, the number of quadruple systems in the survey sample. 
These definitions assume that there is no restriction on magnitude
differences between companions, or on separations between companions, an
assumption that is clearly  unattainable in practice. The quantity that
one can measure is a {\it restricted binary fraction}, which is the binary
fraction restricted  to a stated magnitude difference between primaries
and their secondaries, and in a given (physical, not angular) separation
range. Hopefully, the stated restriction guarantees a {\it complete}
sample, {\it i.e.,} that all binaries within the given separation and
magnitude difference ranges would be detected by the given survey. Such a
complete sample may be defined for a single, given survey, but cannot be
determined {\it a posteriori}, when comparing datasets published by
various authors. The best we can do is to choose restrictions which are
the most likely to result in a complete sample over all the published
studies.

Given all these caveats, our restricted sample from Tables 3 \& 4 will
encompass $\Delta K \sim 3$ and  0\arcsec.1$\le \Delta \Theta\le
1.1$\arcsec~for the $\rho$ Oph cloud core. This sample is restricted to
targets that were searched for multiplicity via high-resolution,
near-infrared techniques {\it only} (e.g., by any of the six surveys of 
Ageorges et al. 1997, Costa et al. 2000, Ghez et al. 1993,  Haisch et al.
2002, Simon et al. 1995, or this work). If we restrict our attention to
the sub-sample of Table 3 searched for multiplicity by only these six
surveys (32 of the 49 systems listed), then among this sub-sample there
are 19 systems with at least one separation in the 0\arcsec.1$\le \Delta
\Theta\le$ 1.1\arcsec~range. 
Of the 32 multiple systems listed in Table 3
that were surveyed by the above-listed six surveys, there remain 13
systems with no separations in the restricted range.  In addition, there
are 48 distinct targets searched by these six surveys, which were found to
be single (listed in Table 4). Thus, the {\it restricted binary fraction}
is 19/(19$+$13$+$48), or 24\% $\pm$ 11\%. We note parenthetically that
when a binary is found in this restricted 0\arcsec.1$\le \Delta \Theta\le$
1.1\arcsec~separation range, if it is part of a wider-separation triple or
quadruple, we still consider the larger multiple system as a single target
for counting purposes.   Were we to count targets that are separated by
$\ge$ 1.1$^{\prime\prime}$ as individual targets for counting purposes, then
we would get a lower value for the restricted binary fraction for $\rho$
Oph. We therefore adopt the value of 24\% $\pm$ 11\%  (19/(32$+$48)) as
the  {\it restricted binary fraction}, $F_{b,r}$, for the $\rho$ Ophiuchi
pre-main-sequence population in the $\sim$ 0\arcsec.1--$\sim$ 1.1\arcsec,
$\Delta K \le 3$ range.  

Ghez et al. (1993) define their {\it complete sample} by the restrictions
$\Delta K \le 2$, 0\arcsec.1 $\le \Theta \le$ 1\arcsec.8 for Taurus and
$\Delta K \le 2$, 0\arcsec.1 $\le \Theta \le$ 1\arcsec.6 for Sco-Oph,
since they assumed distances of 140 pc and 160 pc to Taurus and Ophiuchus,
respectively, and the restriction on angular separation was meant as a
restriction on true, projected, {\it physical} separations, corresponding
to 16--252 AU in that study. The restricted binary star frequency among
Oph-Sco targets,  with the restrictions as defined by by Ghez et al.
(1993) for Sco-Oph was 29\% $\pm$ 12\%  from a sample of 21 targets. The
restricted binary star fraction for this same set of constraints for
Taurus was found to be 37\% $\pm$ 9\% from the 43 targets in their
complete sample. Our newly derived value of 24\% $\pm$ 11\%
from 80 Ophiuchus-only targets, restricted to a somewhat smaller physical separation range, but to a similar
magnitude difference range between primaries and secondaries, is
consistent, within the errors, with the restricted binary star frequency among the pre-main-sequence populations of Ophiuchus and Taurus being identical.

In a related study of X-ray selected T-Tauri stars in the Sco-Cen OB
Association, which is at the same distance as the Ophiuchus cloud core, 
K\"ohler et al. (2000) used both near-infrared  speckle and near-infrared
direct imaging techniques in order achieve completeness  limits for
companions in the 0\arcsec.13  $\le a \le$ 6\arcsec and $\Delta K \le 2.5$
mag ranges.  They found 27 binaries and 2 triples from 88 targets,
corresponding to 33\% $\pm$ 11\% for their restricted binary fraction,
$F_{b,r}\ =\ (27\ +\ 2)/88)$, and 35\% $\pm$ 11\% for their restricted
companion fraction, $F_{c,r}\ =\ (27\ +\ 2\ \times\ 2)/88)$, after
correcting for contamination by background objects and for  X-ray
selection bias.  Although at first glance the restricted binary star
frequency among the Sco-Cen and Ophiuchus samples is consistent with being
identical within the errors, one must bear in mind that the separation
range of $2\arcsec \le a \le 6\arcsec$ was not sampled in Ophiuchus, whereas
it was sampled in Sco-Cen. 

We note that the physical separation range to which our discussion of the
Ophiuchus and Taurus samples was restricted, would correspond to an
angular separation range of 0\arcsec.03 $\le a \le$ 0.37\arcsec~ at the
distance to the Orion star-forming region.  This angular separation range,
although accessible with the largest (10-meter diameter) ground-based
telescopes equipped with adaptive optics, has not yet been explored
towards Orion \cite{be03, sim99}. Therefore, at present, no direct
comparison can be made between the restricted binary fractions in $\rho$
Oph and Taurus, on the one hand, and Orion, on the other.

Comparison of the $\rho$ Oph multiplicity fraction with that of the Main
Sequence requires calculation of the {\it companion star fraction}, $F_c$,
in this separation range. For the same restrictions ($\sim$
0\arcsec.1--$\sim$ 1.1\arcsec, $\Delta K \le 3$), as above, the restricted
companion star fraction derived from the data presented  in  Tables 3 \&
4, is 24\% $\pm$ 11\% for $\rho$ Oph.  This fraction is arrived at  by
counting three triples, V853, ROXs 42C, and ROXs 47A,  from Table 3 as
binaries, since the restriction on angular separations excludes some of
the multiple components that are detected in these systems. The Main
Sequence {\it companion star fraction} in the projected physical
separation range, 16-252 AU, (corresponding to  angular separations
0\arcsec.1 $\le \Theta \le$ 1\arcsec.8 at an assumed 140 pc distance to
$\rho$ Oph) is 16\% $\pm$ 3\%  for stars in the 0.8 M${\odot} \le M \le$
1.3 M$_{\odot}$ mass range (Duquennoy \& Mayor 1991), and 12\% $\pm$ 4\%
for lower mass M dwarfs (Fischer \& Marcy 1992). For purposes of
comparison with the Main Sequence results, it must be borne in mind that 
the restricted companion star fraction derived for $\rho$ Oph above is, a
strict {\it lower} limit in the sense that we imposed an additional
constraint, that of $\Delta K \le 3$ on our sample, whereas no component
flux ratio limits were imposed on the samples used to derive the Main
Sequence restricted companion fractions. An over-abundance of multiple
systems among the pre-main-sequence population of Ophiuchus relative to
the Main Sequence  over this restricted separation range is not yet a
statistically significant (3 $\sigma$) result, however.

Future multiplicity surveys of the $\rho$ Oph pre-main-sequence population
at larger component separations, when combined with the data presented
here, would decide the question of whether we are observing a true
over-abundance of multiple systems among recently formed stars relative to
the Main Sequence, or if we are witnessing the evolution of binary
separations with time.  

\subsection{IRS 43, IRS 44, and IRS 51}

As alluded to in the previous section, often several sets of authors have
searched the same target for multiplicity (viz., Table 4). In general,
search results reported by different sets of authors, but obtained with
techniques sensitive to similar magnitude differences and to similar
size-scales, agree well.  However, there are three objects,  IRS 43, IRS
44, and IRS 51, for which results amongst  different authors, and even
among the same authors, give conflicting or confusing results.

IRS 43 (YLW15A) is an unusual object in $\rho$ Oph, because it is by far
the strongest X-ray source, exhibiting powerful X-ray flaring activity
(Grosso et al. 1997;  Montmerle et al. 2000; Tsuboi et al. 2000). A VLA
3.6cm radio map of IRS 43 resolved two sources: VLA 1, a resolved, thermal
jet, and VLA 2, a point source, positionally coincident (to within $\pm$
0\arcsec.2) with the bright near-infrared source, IRS 43 (Girart,
Rodriguez, \& Curiel 2000). Ground-based, mid-infrared imaging at the Keck
II telescope with JPL's MIRLIN camera found a 0\arcsec.51, PA$=$
332.7$^{\circ}$ binary associated with IRS 43, similar in appearance to
the double 3.6 cm source (Haisch  {\it et al.} 2002). However,
ground-based near-IR measurements detected only a single NIR source (to
K$<$12),  associated with VLA 2,the South-Eastern 10$\mu$m component
(Simon et al. 1995; Costa et al. 2000).  IRS 43 was reported as a single
source with {\it NICMOS}  imaging through the 1.1 $\mu$m and 1.6 $\mu$m
filters (Allen {\it et al.} 2002), but ``two point-like sources clearly
showing the {\it NICMOS} diffraction pattern''  are reported by Terebey
{\it et al.} 2001, who reached a 2000:1 dynamic range using the 2.05
$\mu$m filter.  Nevertheless, Terebey et al. (2001) classify IRS 43 as a
single object, with no further information given.

IRS 44 (YLW 16A) is, similarly, not resolved by ground-based measurements
(Simon et al. 1995; Costa et al. 2000; Haisch et al. 2002). Allen {\it et
al.} (2002) detect two non-point sources, separated by 0\arcsec.5 at 
PA$=$270$^{\circ}$ with a flux ratio of 1.5 at 1.1 $\mu$m and 1.1 at 1.6
$\mu$m in  their {\it NICMOS} images of IRS 44. They interpret this as
scattered light  from possibly a star/disk/envelope system. Terebey {\it
et al.} (2001), on the other hand, report the detection of two point
sources, at a separation of 0\arcsec.27 at P.A. $=$ 81$^{\circ}$, with the
primary detected in all three  (F160W, F187W, and F205W) {\it NICMOS}
images, and the secondary detected only  at the longest, 2.05 $\mu$m,
wavelength.

IRS 51, although found to be a point source by Simon et al. (1995) and by
Costa et al. (2000), was found to be extended in H, K, and L shift-and-add
imaging by Haisch et al. (2002). Whereas the near-infrared images
presented by Haisch et al. are consistent with a secondary at P.A. $=$
10$^{\circ}$ at 1\arcsec.5 separation, the mid-infrared image of the same
object, at similarly high angular resolution, consists of extended
emission, with an emission ``knot'' at just 0\arcsec.7 separation at P.A.
$=$ 15$^{\circ}$. Clearly, the source of the extended near- and
mid-infrared emission surrounding this object deserves further study.

\section{Summary}

\begin{itemize}

\item We have carried out a new, near-infrared, speckle imaging survey of
19 pre-main-sequence objects in $\rho$ Ophiuchus, of which four are
binary, and one is a newly discovered triple system.

\item We have tabulated all the close binaries known in the $\rho$ Oph
cloud core, as well as all of the $\rho$ Oph cloud members searched for
multiplicity, but found to be single, by various high-angular resolution,
near-infrared techniques.

\item Synthesis of the multiplicity data presented here results in the
determination of 24\% $\pm$ 11\% for the {\it restricted binary fraction}
of the Ophiuchus pre-main-sequence population in the  $\sim$
0\arcsec.1--1.1\arcsec, $\Delta K \le 3$ range. This can be 
compared with the 37\% $\pm$ 9\% {\it restricted binary fraction} found in
the Taurus pre-main-sequence population over a $\Delta K \le 2$,
0\arcsec.1 $\le \Theta \le$ 1\arcsec.8 range. Future adaptive optics
observations towards Orion with 10-meter aperture telescopes,  which can
sample the 0\arcsec.03 $\le$ 0.4\arcsec.0 angular separation range, will
allow a direct comparison with the {\it restricted binary fraction} in
Orion in the projected physical separation range, 16AU-154 AU.

\item The observed {\it restricted companion fraction} derived from the 
data presented here is also 24\% $\pm$ 11\% for $\rho$ Oph in a range
restricted to $\sim$ 0\arcsec.1--1.1\arcsec separations, and to
magnitude differences,  $\Delta K \le 3$. Due to the magnitude difference
constraint imposed on the $\rho$ Oph sample,  24\% $\pm$ 11\% represents a
strict {\it lower limit} to the true {\it restricted companion fraction} of
$\rho$ Oph in this separation range. In the physical
separation range of 16-252 AU, the {\it true restricted companion
fraction} for the Main Sequence is 16\% $\pm$ 3\% for stars in the 0.8
M$_{\odot} \le M \le$ 1.3 M$_{\odot}$ mass range (Duquennoy \& Mayor
1991), and 12\% $\pm$ 4\% for lower mass, Main Sequence M dwarfs (Fischer
\& Marcy 1992).

\item Larger surveys for companions among the $\rho$ Oph association
members are required to definitively establish (to 3$\sigma$) the
over-abundance of the companion fraction relative to that of the Main
Sequence. Surveys for multiples at large separations in the $\rho$ Oph
population, corrected for background contamination, will help improve the 
statistics. Synthesis of multiplicity data from future surveys at  all
separations will decide whether it's the multiplicity fraction or the 
binary separation distribution that evolves with time from the
pre-main-sequence to the Main Sequence.

\end{itemize} 

\acknowledgments

We would like to thank the Palomar Office and staff for their help,
expertise, and professionalism  in running the Palomar Observatory. This
research has made use of  NASA's Astrophysics Data System Bibliographic
Services, the SIMBAD database, operated at CDS, Strasbourg, France, and of
the NASA/IPAC Infrared Science Archive, which is operated by the Jet
Propulsion Laboratory, California Institute of Technology, under contract
with the National Aeronautics and Space Administration. MB would
especially like to acknowlege NSF grant AST-0206146 for making her
contributions to this work possible. Additional support for this work
was provided by the National Aeronautics and Space Administrations through
Chandra Award Number AR1-2005A and AR1-2005B issued by the Chandra X-Ray
Observatory Center, which is operated by the Smithsonian Astrophysical
Observatory for and on behalf of NASA under contract NAS8-39073.
\clearpage

\clearpage
\bigskip
\begin{figure}
\figurenum{1}
\plotone{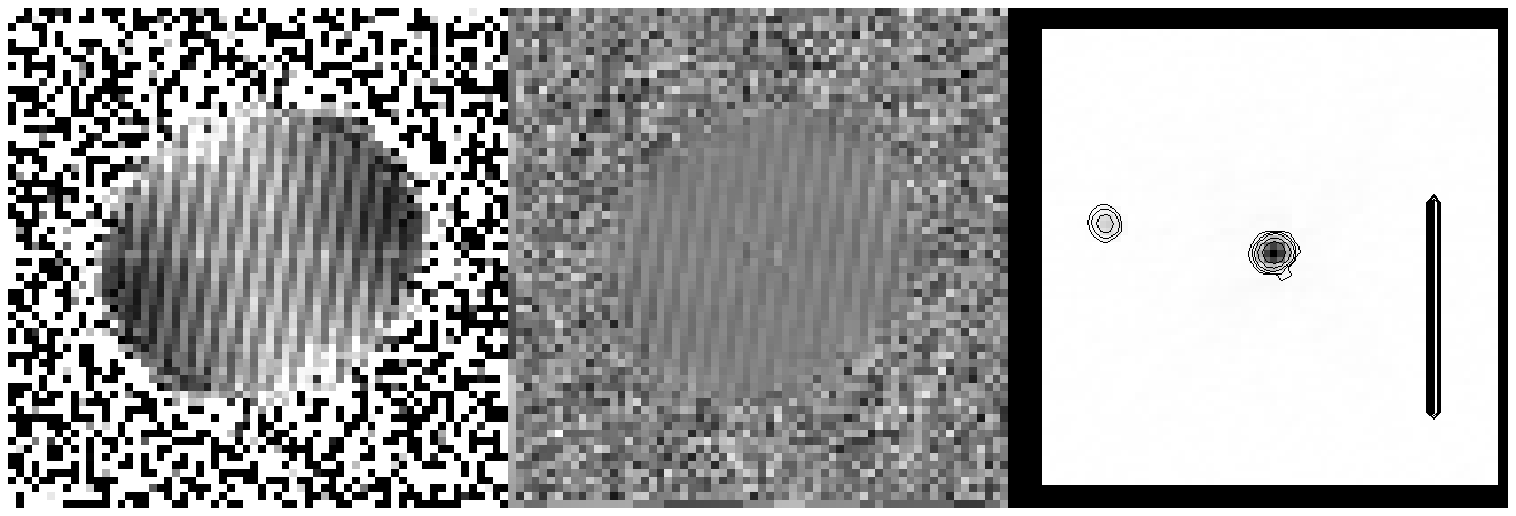}
\caption{From left to right are shown 
the calibrated Fourier power spectrum, Fourier phases,
and the reconstructed image for the newly discovered
hierarchical triple system, ROXs~47A. The Fourier power
spectra and phases exhibit a pattern of narrow stripes 
characteristic of a well-resolved binary.  The superposed single broad stripe
in the power spectrum signals the presence of the third, marginally-resolved
star in the system.  Careful comparison of the
shapes of the stellar images in the reconstructed image
reveals a slight extension in the primary
along a direction perpendicular to the broad bar in the power spectrum.
The simplest explanation for this is that the primary is itself a
binary at 0\arcsec.04 separation. The well-separated third component at
P.A.$=$ 80.8$^{\circ}$ is 0\arcsec.79 away.  The vertical scalebar's length
is 1$^{\prime\prime}$, and the image orientation has North up and East
to the left.}
\end{figure}
\clearpage
\bigskip
\begin{figure}
\figurenum{2}
\plotone{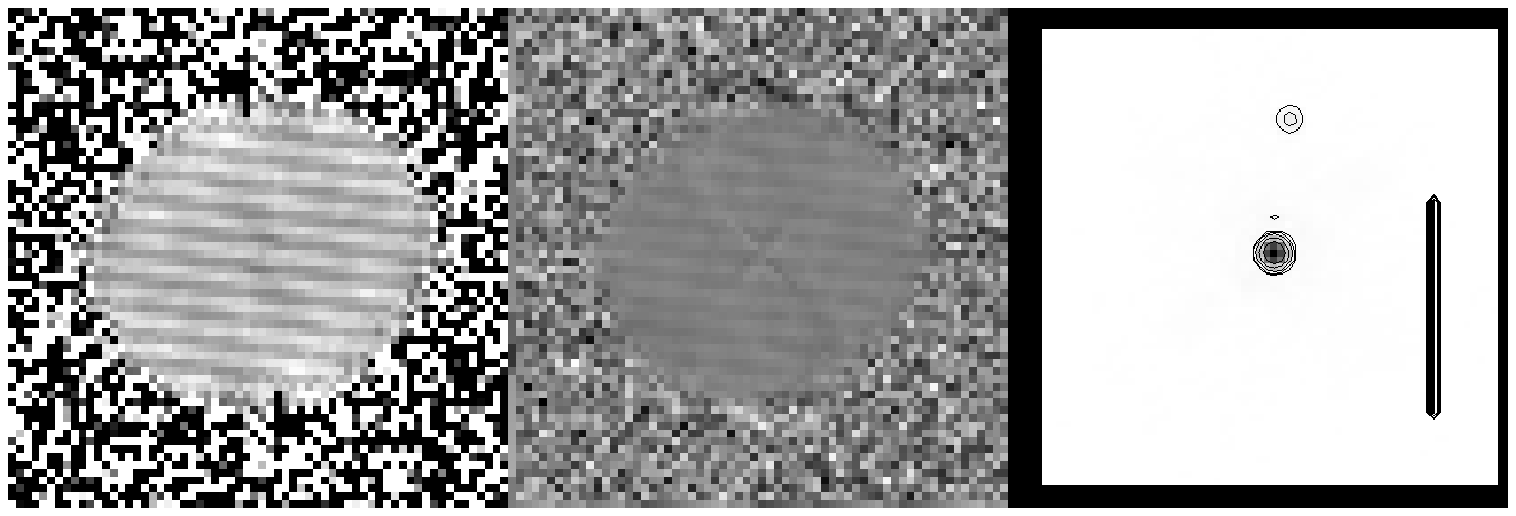}
\caption{From left to right are shown 
the calibrated Fourier power spectrum, Fourier phases,
and the reconstructed image for the ROXs~29 binary sytem. 
The image orientation has North up and East to the left. The vertical
scalebar's length is 1\arcsec.
The binary separation is 0\arcsec.63 at PA 353$^{\circ}$, with a 
brightness ratio of 12.}
%
\end{figure}
\clearpage
\bigskip
\begin{figure}
\figurenum{3}
\plotone{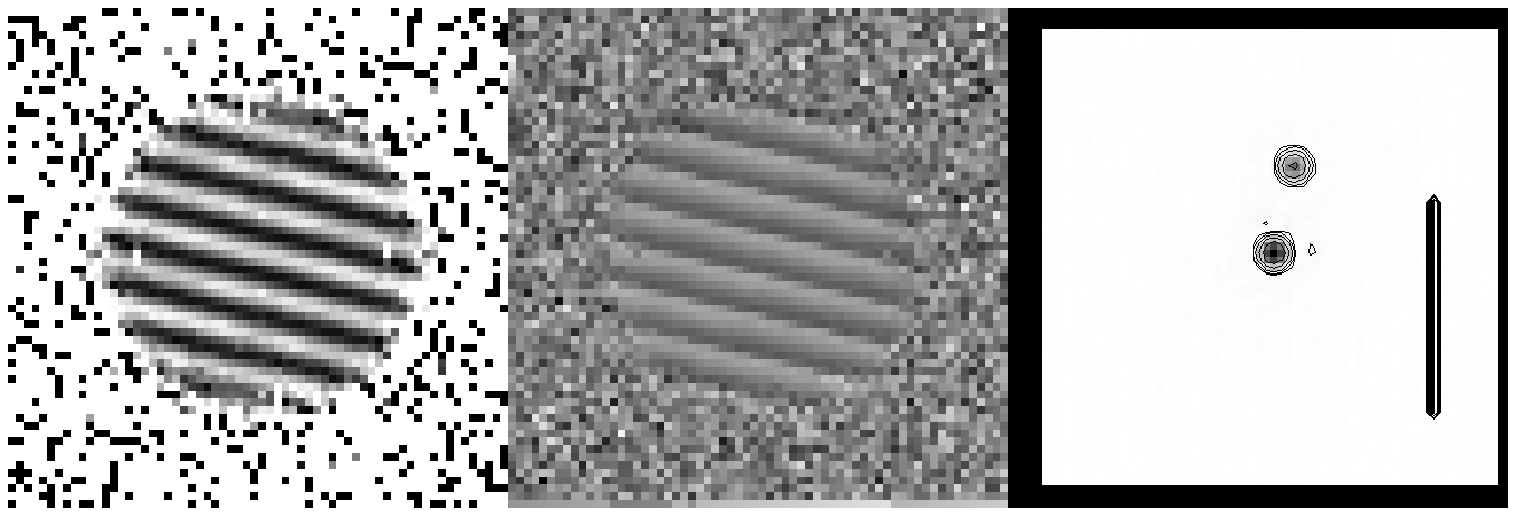}
\caption{From left to right are shown 
the calibrated Fourier power spectrum, Fourier phases,
and the reconstructed image for the ROXs~2 binary system.  
The image orientation has North up and East to the left. The vertical
scalebar's length is 1\arcsec.
The binary separation is 0\arcsec.42 at PA 347$^{\circ}$, with a
brightness ratio of 1.75.}
\end{figure}
\clearpage
\bigskip
\begin{figure}
\figurenum{4}
\plotone{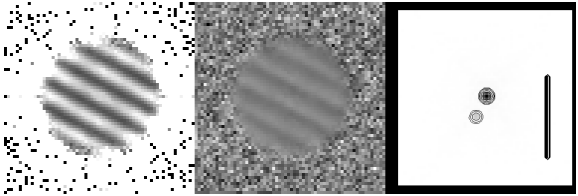}
\caption{From left to right are shown 
the calibrated Fourier power spectrum, Fourier phases,
and the reconstructed image for the ROXs~42C hierarchical triple system.  
The image orientation has North up and East to the left. The vertical
scalebar's length is 1\arcsec.
The source separation is 0\arcsec.28 at PA 153$^{\circ}$, and the
brightness ratio of the resolved components is 4.4. One of these components 
is itself a spectroscopic binary (Mathieu et al. 1989). This is the only system in which a significant change in the orbital parameters was observed. For comparison, the previously
published binary separation was 0\arcsec.16 at PA 135$^{\circ}$
(Ghez et al. 1993).}
\end{figure}
\clearpage
\bigskip
\begin{figure}
\figurenum{5}
\plotone{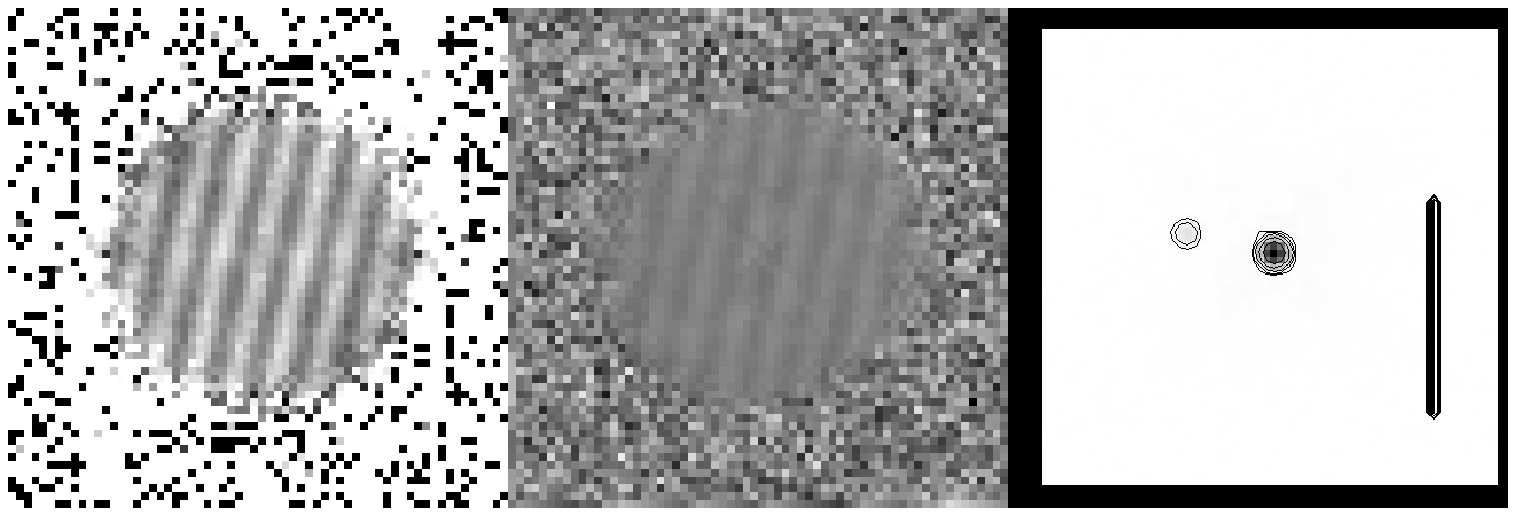}
\caption{From left to right are shown 
the calibrated Fourier power spectrum, Fourier phases,
and the reconstructed image for the IRS~2 binary system.  
The image orientation has North up and East to the left. The vertical
scalebar's length is 1\arcsec.
The binary separation is 0\arcsec.42 at PA 78$^{\circ}$, with a
brightness ratio of 8.}
\end{figure}
\clearpage
\bigskip
\begin{figure}
\figurenum{6}
\plotone{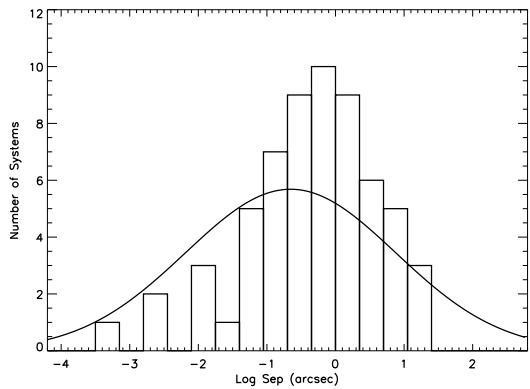}
\caption{Distribution of the separations of the pairs in the
binary and multiple systems of Table 3.  The overplotted curve corresponds
to the distribution of separations in Solar-type Main Sequence stars as
measured by Duquennoy \& Mayor (1991), and is scaled to correspond to the
number of pre-main sequence systems plotted.  The separations for the Main
Sequence stars are computed according to the same assumptions as used by
previous workers: a distance of 140 pc, stellar component masses of
0.5~$M_\odot$, and circular orbits.}
\end{figure}
\clearpage
%
%
%
%
%
%
%
%
%
%
%
%
\input{Table1.tex}
\input{Table2.tex}
\input{Table3.tex}
\input{Table4.tex}
\end{document}

%% file: Table1.tex
\begin{deluxetable}{llcll}
\footnotesize
\tablecaption{ Observing Log} 
\tablewidth{0pt}
\tablehead{
\colhead{Target} & \colhead{Night} & \colhead{Series} & \colhead{RA (2000)} & \colhead{Dec (2000)}}
\startdata
V852 Oph         & 24 May 2002   & 8   & 16 25 22.16 & -24 30 13.7 \\
ROXs 2            & 24 May 2002   & 8   & 16 25 24.37 & -23 55 09.9 \\
IRS2             & 24 May 2002   & 8   & 16 25 36.75 & -24 15 42.1 \\
ROXs 4 IRS10      & 24 May 2002   & 8   & 16 25 50.53 & -24 39 14.3 \\
ROXs 10A/DoAr24   & 24 May 2002   & 8   & 16 26 17.09 & -24 20 21.4 \\
ROXs 9A           & 24 May 2002   & 8   & 16 26 23.99 & -25 47 16.1 \\
ROXs 9B           & 24 May 2002   & 8   & 16 26 30.72 & -25 43 39.8 \\
ROXs 9C           & 24 May 2002   & 8   & 16 26 37.95 & -25 45 12.8 \\
GY292            & 24 May 2002   & 10  & 16 27 33.11 & -24 41 15.1 \\
ROXs 35A          & 24 May 2002   & 8   & 16 27 38.31 & -23 57 32.9 \\
WSB52/GY314            & 24 May 2002   & 8   & 16 27 39.43 & -24 39 15.5 \\
ROXs 29/SR 9       & 24 May 2002   & 8   & 16 27 40.28 & -24 22 04.3 \\
ROXs 40           & 24 May 2002   & 8   & 16 30 51.51 & -24 11 34.0 \\
WSB69A           & 25 May 2002   & 10  & 16 31 04.37 & -24 04 33.3 \\
WSB69b           & 25 May 2002   & 8   & 16 31 05.17 & -24 04 40.3 \\
ROXs 42C          & 24 May 2002   & 8   & 16 31 15.75 & -24 34 02.2 \\
ROXs 44           & 24 May 2002   & 8   & 16 31 33.45 & -24 27 37.1 \\
IRAS 16289-2457   & 24 May 2002   & 8   & 16 31 54.74 & -25 03 23.8 \\
ROXs 47A          & 24 May 2002   & 8   & 16 32 11.81 & -24 40 21.3 \\
\enddata
\end{deluxetable}

%% file: Table2.tex
\begin{deluxetable}{llllr}
\tabletypesize{\footnotesize}
\tablecaption{Characteristics of Multiple Systems Detected in This Survey} 
\tablewidth{0pt}
\tablehead{
\colhead{System} & \colhead{Separation (\arcsec)} & \colhead{P.A. ($^{\circ}$)}  & \colhead{Flux Ratio\tablenotemark{1}} & \colhead{Ref.}
}
\startdata
ROXs 2                     & 0.42  $\pm$ 0.03    & 347.1 $\pm$ 0.2    & 1.75 $\pm$ 0.1             & This work\\
                           & 0.41  $\pm$ 0.03    & 349   $\pm$ 1      & 2.58 $\pm$ 0.3             & C00 \\
IRS2                       & 0.42  $\pm$ 0.03    & 77.6  $\pm$ 0.4    & 8.0  $\pm$ 2.5             & This work \\
                           & 0.44  $\pm$ 0.03    & 79    $\pm$ 4      & 7.58 $\pm$ 0.4 at H-band   & C00 \\
ROXs 29/SR 9/Elias 34      & 0.63  $\pm$ 0.03    & 353.3 $\pm$ 0.05   & 12   $\pm$ 5               & This work \\
                           & 0.59  $\pm$ 0.01    & 350   $\pm$ 1      & 11   $\pm$ 2; 13$\pm$2     & G93 \\
ROXs 42C\tablenotemark{2  }& 0.28  $\pm$ 0.03    & 152.9 $\pm$ 0.5    & 4.4  $\pm$ 0.7             & This work\\
                           & 0.157 $\pm$ 0.003   & 135   $\pm$ 3      & 4.0  $\pm$ 0.34            & G93\\
ROXs 47A                   & 0.79  $\pm$ 0.03    &  80.8 $\pm$ 0.1    & 2.5  $\pm$0.2              & This work  \\
ROXs 47A\tablenotemark{3  }& 0.04  $\pm$ 0.03    & 107   $\pm$ 20     & 1.1  $\pm$0.7              & This work \\
\enddata
\tablenotetext{1}{All flux ratios are at K-band, except where otherwise noted.}
\tablenotetext{2}{This is a triple system, with one component consisting of\\
 a spectroscopic binary (Mathieu et al. 1989)}.
\tablenotetext{3}{Due to the extreme proximity of the components, there remains a 180$^{\circ}$\\
ambiguity in the position angle of the close binary in this hierarchical triple system.}
\tablerefs{C00$=$Costa et al. 2000; G93$=$Ghez et al. 1993}
\end{deluxetable}

%% file: Table3.tex
\clearpage
\begin{deluxetable}{lllrccrl}
\rotate
\tabletypesize{\scriptsize}
\tablewidth{0pt}
\tablecolumns{8}
\tablecaption{Separations of Binary/Multiple Systems in Ophiuchus}
\tablehead{
\colhead{Name} &
 \colhead{$\alpha$(2000)} &
 \colhead{$\delta$(2000)} &
 \colhead{System} &
 \colhead{Sep'n.} & 
 \colhead{P.A.}   & 
 \colhead{Refs.\tablenotemark{3}} &
 \colhead{Other}\\
 \colhead{}     & 
 \colhead{}     & 
 \colhead{}     & 
 \colhead{Type\tablenotemark{1}} & 
 \colhead{Arcsec}  & 
 \colhead{Deg\tablenotemark{2}} & 
 \colhead{}                      &
 \colhead{Aliases}}
\startdata
155203-2338     &15 54 59.87&-23 47 18.2&B &0.80        &229    &G93            &               \\
WSB 3           &16 18 49.61&-26 32 53.3&B &0.60        &162    &RZ93           &               \\
WSB 4           &16 18 49.66&-26 10 06.2&B &2.84        & 128.5 &K02,RZ93       & 
 IRC\tablenotemark{4} (RZ93)        \\
WSB 11\tablenotemark{5}&16 21 57.3&-22 38 16&B&0.50     &N.A.   &RZ93       &                   \\
WSB 18          &16 24 59.79&-24 56 00.3    &T(Q?) &1.08&80.4   &K02,RZ93   &Faint 2MASS src. @ 10" sepn\\
WSB18A          &           &               &      &0.100&339.55&K02,RZ93   &Primary is double \\
WSB 19          &16 25 02.13&-24 59 31.8    &B     &1.53 &260.7 &K02,RZ93   &           \\
WSB 20\tablenotemark{5} &16 25 10.5 &-23 19 14&B    &1.0 & 23   & GM01,RZ93 &   \\
ROX 1 &16 25 19.28&-24 26 52.1&B   &0.236 &156 &G93,A32    &SR2, SAO 184375, Elias 6, GSS 5, 162218-2420  \\
HD 147889\tablenotemark{5}&16 25 24.32&-24 27 56.6&B&4.09$\times$10$^{-4}$&N.A.&HM95&SpB; Period=5d; SR1, Elias 9, GSS9 \\
ROXs 2           &16 25 24.37&-23 55 09.9    &B     &0.41 &349   &C00;This work&           \\
IRS 2           &16 25 36.75&-24 15 42.1    &B     &0.44+/-0.03 &79+/-4 &C00;This work&   \\
ROXs 5           &16 25 55.86&-23 55 10.1    &B     &0.13        &130    &A97          &    \\     
WSB 26          &16 26 18.40&-25 20 55.6    &B     &1.15        & 23.8  &K02,RZ93     & DoAr 23   \\
WSB 28          &           &               &B     &5.1         &358    &RZ93         &            \\
WSB 28 Primary  &16 26 20.99&-24 08 51.8    &      &            &       &             &            \\
WSB 28 Secondary&16 26 20.99&-24 08 46.7    &      &            &       &             &            \\
DoAr 24E&16 26 23.38&-24 20 59.7&B &2.05 &148.6 & K02,C00,A97,S95,C88 & ROXs 10B, Elias 22, GSS31; 
 IRC\tablenotemark{4} \\
Elias23$+$GY21    &            &              &B &10.47&322.6  & HBGR02                &             \\
Elias 23        &16 26 24.06 &-24 24 48.1   &  &     &       & GSS32 is single       &S2,GSS32,GY23 \\
                &            &              &  &     &       & (S95;T01;HBGR02) &              \\
GY 21           &16 26 23.60 &-24 24 39.4   &  &     &       & GY21 is single  &              \\
                &            &              &  &     &       & (C00;HBGR02)&              \\
VSSG 27         &16 26 30.50&-24 22 57.1    &B &1.22+/-0.03  &68 +/-1 &C00           &GY51           \\
S1       &16 26 34.18&-24 23 28.2    &B &0.020        &110     &S95          &ROXs 14, GY70, GSS35, Elias 25 \\
WSB 35\tablenotemark{5}&16 26 34.8 &-23 45 41      &B &2.29 &130.3  & K02,RZ93   &DoAr 26        \\
GSS37   &16 26 42.87&-24 20 29.8    &B(T?) &1.44  & 67.0  &K02,C00,RZ93&ROX 15, Elias 26;2MASS source @ 9.3"\\
VSS27   &16 26 46.44&-24 12 00.0    &B     &0.59     &104.6  &A97,C00     & ROXs 16, WSB38       \\
WL 2            &           &               &B     &4.17     &343    & B89        &              \\
WL 2(A)         &16 26 48.50& -24 28 38.7  & &       &       &            & \\
WL 2(B)         &16 26 48.42& -24 28 34.7  & &       &       &            & \\
WL 18           &16 26 48.99& -24 38 25.1  &B        &3.55 & 293   & B89&       \\
VSSG 3    &16 26 49.25  &-24 20 02.9 &B &0.25+/-0.03&47 +/-1& C00 &GY135; @24.5" is BKLT 162648-241942\\
SR 24           &           &           &T &6.00(5.093")& 60    &S95;(2MASS)    & WSB 42 \\
SR 24A    &16 26 58.52&-24 45 36.7&  &            &       &SR 24S         &G93 found SR 24S to be single\\
SR 24B          &16 26 58.45&-24 45 31.7&  &0.197       & 84    &S95;C00        & SR 24N, DoAr 29  \\ 
WL 1            &16 27 04.12 & -24 28 29.9 &B        &0.82 & 321.2 & HBGR02;C00 &  \\
SR21     &           &           &B &6.7 (6.33") &175    &S95;(2MASS)    & Elias 30; VSSG23     \\
SR 21A          &16 27 10.28&-24 19 12.6&  &            &       &               & \\
SR 21B          &16 27 10.33&-24 19 18.9&  &            &       &               & \\
ROXs 20A$+$ROXs 20B  &           &           &B&10.0    &126    &BA92        &              \\
ROXs 20A         &16 27 14.51&-24 51 33.4& &        &       & Secondary  & WSB45, HBC640  \\
ROXs 20B         &16 27 15.15&-24 51 38.8& &        &       & Primary    & WSB46, HBC641  \\
WL20            &           &           &T&        &       &RB01        &               \\
WL20W           &16 27 15.69&-24 38 43.4& & 3.17   &270    &RB01        &E-W sep'n.    \\ 
WL20S           &16 27 15.72&-24 38 45.6& & 2.26   &173    &RB01        &S-W sep'n.   \\
WL20E           &16 27 15.89&-24 38 43.4& & 3.66   &232    &RB01        &S-E sep'n. \\ 
SR 12$+$IRS42     &           &            &T(?)     &       &            &This work, BKLT97, 2MASS & \\
SR 12   &16 27 19.52&-24 41 40.4 &B &0.30+/-0.03 &85            & S95;C00            &ROXs 21, IRS40, GY250 \\
IRS42           &16 27 21.48&-24 41 43.0 &         &26.8 & 85.8         & S95;C00 found IRS42 single  & GY252  \\
GY263$+$GY265   &           &            &B(T?) &6.99  &322  &HBGR02;S95 &YLW15A;  CRBR85 is 34" away  \\
GY263           &16 27 26.63&-24 40 44.9 &           &      &             &               & \\
IRS 43\tablenotemark{6}&16 27 26.94&-24 40 50. &&     &      &HBGR02;T01;C00;S95          &GY265 \\
IRS 44$+$GY262           &           &           &T(?) &      &             &              & \\
GY262                  &16 27 26.49&-24 39 23.0&  &23.21&              &T01      & \\
IRS 44\tablenotemark{6}&16 27 28.01&-24 39 33.6&B &0.27  &81           &HBGR02;T01;C00;S95      & YLW16A  \\
WL 13           &16 27 27.40& -24 31 16.6  &B        &0.46 & 356   & C00        &VSSG25; Elias 31 \\
VSSG 17         &16 27 30.17&-24 27 43.5&B  &0.25+/-0.03&26 +/-6& C00           &Elias 33/IRS47/GY279 \\
DoAr32$+$DoAr33   &           &           &T &48          &170    &BA92           &  \\      
DoAr32  &16 27 38.31&-23 57 32.9&B &10.06       &332.1  &A97            &  WSB51, ROXs 30B (Primary)\\ 
DoAr33  &16 27 39.00&-23 58 19.0&&          &       &                 &  WSB53, ROXs 30C  (Secondary)\\
SR9      &16 27 40.28& -24 22 04.3&B         &0.59   & 350& GM01,G93;This work & ROXs 29, Elias 34, GY319\\
VSSG 14         &16 27 49.86&-24 25 40.5&B  &0.101      &89     & S95 &Elias 36; GY372 is 18.34" away \\
ROXs 31          &16 27 52.07&-24 40 50.4&B &0.39        &71.6   &A97,S95,S87,C00& IRS 55/GY380/HBC642\\
SR 20        &16 28 32.68&-24 22 45.0&B &0.071   &225    &G93;S95        & ROX 33, WSB61, HBC 643; DoAr 38\\
V853 Oph     &16 28 45.29&-24 28 18.9 &T &0.43        &97     &S95;G93;C00;GM01    & SR 13, WSB62   \\
V853 Oph A    &           &            &  &0.013       &96     &S95            & Primary resolved  \\ 
Haro1-14c$+$Haro1-14&         &           &B&12.9    &122    &RZ93         &                \\
Haro1-14c       &16 31 04.37&-24 04 33.3& &        &       &             & Primary        \\
Haro1-14        &16 31 05.17&-24 04 40.3& &        &       &             & WSB69; Secondary      \\
ROX 42B         &16 31 15.02& -24 32 43.7&B&0.056       &89     &S95  &           \\
ROXs 42C         &16 31 15.75&-24 34 02.2&T &0.157       &135    &G93;This work+SpB M89  &   \\
ROXs 42C         &           &           &  & 0.00195    &       & SpB P=35.95 days; M89 & \\
ROXs 43          &        &       &Q &4.67/4.80/4.3(4.39)&13.7/7/13 &A97/S95/RZ93(2MASS)&A and B separation \\
ROXs 43 A        &16 31 20.12&-24 30 05.0&  & 0.0028     &N.A.       &M89     & SpB P=89.1 $\pm$ 0.2 days\\   
ROXs 43 B        &16 31 20.19&-24 30 00.7&          &0.016       &89     &S95     &                \\
WSB 71          &           &      &B(T?)&3.56 & 35.0  & K02;S95 & Faint 2MASS source 7.330" from WSB 71A\\
WSB 71A         &16 31 30.88&-24 24 39.9    &     &     &       &         &      \\
WSB 71B         &16 31 31.04&-24 24 37.0    &     &     &       &         &      \\
L1689 IRS5      &16 31 52.12& -24 56 15.7  &B        &2.92 & 240.3 & HBGR02     &L1689 SNO2 \\
ROXs 45E$+$ROXs 45F &           &          &B&15.0    &75    &BA92         &                \\
ROXs 45E         &16 32 00.5 &-25 30 29 & &        &       &             & DoAr49       \\
ROXs 45F         &16 32 01.6 &-25 30 25 & &        &       &             & DoAr50       \\
ROXs 47A         &16 32 11.81&-24 40 21.3&T        &0.813  &80.5         &This work &     \\ 
ROXs 47Ab        &           &           &B        &0.046  &107.9 or 287.9&This work &    \\
HD 150193       &16 40 17.93&-23 53 45.3 &B(T?)  &1.09 &221.7& K02  & Elias 49 (HAeBe); 2MASS src. @ 13.1"  \\
%
%
\enddata
\tablenotetext{1}{B$=$Binary; SpB$=$Spectroscopic Binary; T$=$Triple; Q$=$Quadruple}
\tablenotetext{2}{PAs are E of N, measured from the primary at K}
\tablenotetext{3}{Separations and PAs listed are from the first reference}
\tablenotetext{4}{IRC$=$Infrared Companion system}
\tablenotetext{5}{Coordinates for WSB 11, WSB 20, \& WSB 35 are from the SIMBAD database; \\
these sources fell outside the 2MASS survey's areal coverage.}
\tablenotetext{6}{See text for discussion of IRS 43, IRS 44, \& IRS 51.}
\tablerefs{A97$=$Ageorges et al. 1997; A32$=$Aitken 1932; B89$=$Barsony {\it et al.} 1989;\\
BKLT97$=$Barsony et al. 1997; BA92$=$Bouvier \& Appenzeller 1992; C88$=$Chelli et al. 1988;\\  
C00$=$Costa et al. 2000; GM01$=$Geoffray \& Monin 2001; G93$=$Ghez et al. 1993;\\
HM95$=$Haffner \& Meyer 1995; HBGR02$=$Haisch et al. 2002; K02$=$Koresko 2002\\
M89$=$Mathieu {\it et al.} 1989; RZ93$=$Reipurth \& Zinnecker 1993; S95$=$Simon et al. 1995;\\
T01$=$Terebey et al. 2001}
\end{deluxetable}

%% file: Table4.tex
\begin{deluxetable}{llll}
\clearpage
\tabletypesize{\footnotesize}
\tablewidth{0pt}
\tablecolumns{4}
\tablecaption{Single Targets Previously Searched for Multiplicity} 
\tablehead{
\colhead{Object Name} &  \colhead{RA (2000)} & \colhead{Dec (2000)} &\colhead{Reference}}
\startdata
WSB 16/DoAr 15\tablenotemark{1}& 16 23 34.7   & -23 40 29.0     & A97 \\
V852 Oph              & 16 25 22.16           & -24 30 13.7     & This work\\
IRS 9                 & 16 25 49.07           & -24 31 38.8     & C00 \\
ROXs 3                & 16 25 49.65           & -24 51 31.7     & S95 \\
ROXs 4/IRS10          & 16 25 50.53           & -24 39 14.3     & This work\\
ROXs 6/SR 4           & 16 25 56.18           & -24 20 48.2     & G93;S95;C00 \\
ROXs 7/IRS 13/GSS20\tablenotemark{2}& 16 25 57.54& -24 30 31.7  & A97; C00; IRS11 is 25.705\arcsec away\\
ROXs 8/DoAr 21/GSS23  & 16 26 03.05           & -24 23 01.63    & G93;S95;A97;C00\\
GSS25/SR 3            & 16 26 09.33           & -24 34 12.1     & S95; C00 \\
GSS26                 & 16 26 10.35           & -24 20 54.7     & C00 \\
GSS29                 & 16 26 16.87           & -24 22 23.0     & S95 \\
ROXs 10A/DoAr24       & 16 26 17.09           & -24 20 21.4     & G93;This work\\
VSSG1/Elias 20        & 16 26 18.89           & -24 28 19.6     & C00 \\
GSS30 IRS1/Elias 21\tablenotemark{3}& 16 26 21.42& -24 23 02.3  & C00 \\
DoAr25/GY17           & 16 26 23.69           & -24 43 14.0     & C00 \\
ROXs 9A               & 16 26 23.99           & -25 47 16.1     & This work\\
Elias 24/WSB 31       & 16 26 24.09           & -24 16 13.3     & C00 \\
ROXs 9B               & 16 26 30.72           & -25 43 39.8     & This work\\
ROXs 9C               & 16 26 37.95           & -25 45 12.8     & This work\\
WSB 37/GY93           & 16 26 41.28           & -24 40 17.8     & C00 \\
WL16                  & 16 27 02.35           & -24 37 27.2     & S95; C00 \\
WL15/Elias 29         & 16 27 09.44           & -24 37 18.7     & S95; C00 \\
GY224                 & 16 27 11.20           & -24 40 46.6     & HBGR02 \\
WL19                  & 16 27 11.73           & -24 38 31.9     & HBGR02 \\
WL4                   & 16 27 18.50           & -24 29 05.9     & C00 \\
VSSG 22               & 16 27 22.92           & -24 17 57.3     & C00 \\
IRS 48                & 16 27 37.18           & -24 30 35.2     & S95; C00 \\
IRS32b/GY235\tablenotemark{4}& 16 27 13.84    & -24 43 31.6     & S95 \\
GY292                 & 16 27 33.11           & -24 41 15.1     & This work\\
IRS 50                & 16 27 38.12           & -24 30 43.1     & C00 \\
ROXs 35A              & 16 27 38.31           & -23 57 32.9     & This work\\
IRS 49                & 16 27 38.31           & -24 36 58.7     & S95; C00 \\
WSB 52/GY314          & 16 27 39.43           & -24 39 15.5     & S95, This work\\
IRS 51                & 16 27 39.83           & -24 43 14.9     & S95;C00; HBGR02 \\
IRS 56                & 16 27 50.72           & -24 48 21.4     & S95 \\
IRS 54                & 16 27 51.79           & -24 31 45.7     & C00 \\
SR 10                 & 16 27 55.56           & -24 26 18.1     & S95; C00 \\
WSB 60                & 16 28 16.51           & -24 36 58.0     & S95; C00 \\
WSB 63                & 16 28 54.09           & -24 47 44.2     & S95 \\
WSB 67\tablenotemark{5}& 16 30 23.31          & -24 54 16.2     & S95 \\
ROXs 39\tablenotemark{1}&16 30 35.6           & -24 34 15.0     & A97 \\   
ROXs 40               & 16 30 51.51           & -24 11 34.0     & This work\\
WSB 69A               & 16 31 04.37           & -24 04 33.3     & This work\\
WSB 69B               & 16 31 05.17           & -24 04 40.3     & This work\\
ROXs 44/Haro1-16      & 16 31 33.45           & -24 27 37.1     & S95;This work\\
IRS 63                & 16 31 35.67           & -24 01 29.3     & HBGR02 \\
IRAS 16289-2457       & 16 31 54.74           & -25 03 23.8     & This work\\
IRS 67                & 16 32 01.01           & -24 56 41.9     & HBGR02 \\

\enddata
\tablenotetext{1}{Coordinates from SIMBAD database}
\tablenotetext{2}{WSB 16 is associated with extended optical nebulosity (Ageorges et al. 1997)}
\tablenotetext{3}{GSS30 IRS1 coordinates are from Zhang et al. (1997)}
\tablenotetext{4}{Verification of association membership remains to be established (Simon et al. 1995)}
\tablenotetext{5}{Coordinates for WSB 67 are precessed from Simon et al. (1995)}
\end{deluxetable}

%% file: ms.bbl
\begin{thebibliography}{}

\bibitem[Ageorges et al. 1997]{ag97} Ageorges, N., Eckart, A., Monin, J.-L., \& M\'enard, F.
1997, A\& A, 326, 632

\bibitem[Aitken 1932]{ait32} Aitken, R.G. 1932, New General Catalogue of Double Stars 
within 120 degrees of the North Pole, Carnegie Institution of Washington: Washington, D.C. 
(Publication No. 417)

\bibitem[Allen et al. 2002]{all01} Allen, L.E., Myers, P.C., DiFrancesco, J., Mathieu, R.,
Chen, H., \& Young, E. 2002, ApJ, 566, 993

\bibitem[Aspin et al. 1997]{asp97} Aspin, C., Puxley, P.J., Hawarden, T.G., Paterson, M.J.,
\& Pickup, D.A. 1997, MNRAS, 284, 257

\bibitem[Barsony et al. 1989]{bar89} Barsony, M., Burton, M.G., Russell, A.P.G., Carlstrom, J.E.,
\& Garden, R. 1989, ApJ, 346, L93

\bibitem[Barsony et al. 1997]{bar97} Barsony, M., Kenyon, S.J., Lada, E.A., \& Teuben, P.J.
1997, ApJS, 112, 109

\bibitem[Beck et al. 2003]{be03} Beck, T., Simon, M., \& Close, L.M. 2003, ApJ, 583, 358

\bibitem[Bontemps et al. 2001]{bon01} Bontemps, S., Andr\'e, P., Kaas, A.A., Nordh, L., Olofsson, G.,
Huldtgren, M., Abergel, A., Blommaert, J., Boulanger, F., Burgdorf, M., Cesarsky, C.J., Cesarsky, D.,
Copet, E., Davies, J., Falgarone, E., Lagache, G., Montmerle, T., P\'erault, M., Persi, P., Prusti, T.,
Puget, J.L., \& Sibille, F. 2001, A\& A, 372, 173

\bibitem[Bouvier \& Appenzeller 1992]{boa92} Bouvier, J. \& Appenzeller, I. 1992, A\& AS, 92, 481

\bibitem[Casanova et al. 1995]{cas95} Casanova, S., Montmerle, T., Feigelson, E.D., \& Andr\'e, Ph.
1995, ApJ, 439, 752

\bibitem[Chelli et al. 1988]{che88} Chelli, A., Cruz-Gonzalez, I., Zinnecker, H., Carrasco, L., 
\& Perrier, C. 1988, A\& A, 207, 46

\bibitem[Costa et al. 2000]{co00} Costa, A., Jessop, N.E., Yun, J.L., Santos, C., Ward-Thompson, D.,
\& Casali, M.M. 2000 in {\it Birth and Evolution of Binary Stars}, Poster Proceedings of IAU Symp. No. 200,
eds. B. Reipurth \& H. Zinnecker, p. 48

\bibitem[Dolidze \& Arakelyan 1959]{doar95} Dolidze, M.V. \& Arakelyan, M.A. 1959, AZh, 36, 444

\bibitem[Duchene 1999]{duc99} Duchene, G., 1999, A\& A, 341, 547

\bibitem[Duquennoy \& Mayor 1991]{duq91} Duquennoy, A. \& Mayor, M. 1991, A\& A, 248, 485

\bibitem[Fischer \& Marcy 1992]{fis92} Fischer, D.A. \& Marcy, G.W. 1992, ApJ, 396, 178

\bibitem[Geoffray \& Monin 2001]{geo01} Geoffray, H. \& Monin, J.-L. 2001, A\& A, 369, 239

\bibitem[Ghez et al. 1993]{ghe93} Ghez, A., Neugebauer, G., \& Matthews, K. 1993, AJ, 106, 2005

\bibitem[Girart et al. 2000]{gir00} Girart, J.M., Rodriguez, L.F., \& Curiel, S. 2000, ApJ, 544, 153

\bibitem[Grosso et al. 1997]{gro97} Grosso, N., Montmerle, T., Feigelson, E.D., Andr\'e, P.,
Casanova, S., \& Gregorio-Hetem, J. 1997, Nature, 387, 56

\bibitem[Grosso et al. 2000]{gro00} Grosso, N., Montmerle, T., Bontemps, S, Andr\'e, P., \& Feigelson, E.D., 2000, A\& A, 359, 113

\bibitem[Haisch et al. 2002]{hai02} Haisch, K.E., Jr., Barsony, M., Greene, T.P., \& Ressler, M.E. 2002,
AJ, 124, 2841

\bibitem[Haffner \& Meyer 1995]{haf95} Haffner, L.M. \& Meyer, D.M. 1995, ApJ, 453, 450

\bibitem[Hartigan \& Kenyon 2003]{ha03} Hartigan, P. \& Kenyon, S.J. 2003, ApJ, in press

\bibitem[Imanishi et al. 2002]{ima02} Imanishi, K., Koyama, K., \& Tsuboi, Y. 2002, ApJ,, 557 747

\bibitem[Koresko 2002]{kor02} Koresko, C. 2002, AJ, 124, 1082

\bibitem[K\"ohler et al. 2000]{koh00} K\"ohler, R., Kunkel, M., Leinert, C.,
\& Zinnecker, H. 2000, A\& A, 356, 541 

\bibitem[Leinert et al. 1993]{lei93} Leinert, Ch., Zinnecker, H., Weitzel, N.,
Christou, J., Ridgway, S.T.., Jameson, R., Haas, M., \& Lenzen, R. 1993, A\& A, 278, 129 

\bibitem[Mathieu et al. 1989]{mat89} Mathieu, R., Walter, F.M., \& Myers, P.C. 1989, AJ, 987

\bibitem{Lohmann83} Lohmann, A.W., Weigelt, G., \& Wirnitzer, B. 1983, ApOpt, 22, 4028

\bibitem[Montmerle et al. 1983]{mon83} Montmerle, T., Koch-Miramond, L., Falgarone, E., \& Grindlay, J.E.
1983, ApJ, 269, 182

\bibitem[Montmerle et al. 2000]{mon00} Montmerle, T., Grosso, N., Tsuboi, Y., \& Koyama, K. 2000,
ApJ, 532, 1097

\bibitem[Prosser et al. 1994]{pro94} Prosser, C.F., Stauffer, J.R., Hartmann, L.,
Soderblom, D.R., Jones, B.F., Werner, M.F., McCaughrean, M.J. 1994, ApJ, 421, 517

\bibitem[Reipurth \& Zinnecker 1993]{rei93} Reipurth, B. \& Zinnecker, H. 1993, 
A\& A, 278, 81

\bibitem[Ressler \& Barsony 2001]{res01} Ressler, M.E. \& Barsony, M. 2001, AJ, 121, 1098 

\bibitem[Simon et al. 1987]{sim87} Simon, M., Howell, R.R., Longmore, A.J., Wilking, B.A.,
Peterson, D.M., \& Chen, W.-P. 1987, ApJ, 320, 344

\bibitem[Simon et al. 1995]{sim95} Simon, M., Ghez, A., Leinert, Ch., Cassar, L.,
Chen, W.P., Howell, R.R., Jameson, R.F., Matthews, K., Neugebauer, G., \& Richichi, A. 1995, 
ApJ, 443, 625

\bibitem[Simon et al. 1999]{sim99} Simon, M., Close, L.M., \& Beck, T.L. 1999, AJ, 117, 1375

\bibitem[Struve \& Rudjkobing 1949]{sr49} Struve, O. \& Rudjkobing, M. 1949, ApJ, 109, 92

\bibitem[Terebey et al. 2001]{Ter01} Terebey, S., Van Buren, D., Hancock, T., Padgett, D.L.,
\& Brundage, M. 2001, in {\it From Darkness to Light: Origin and Evolution of Young Stellar Clusters},
ASP Conf. Ser., Vol. 243, eds. T. Montmerle \& Ph. Andr\'e, p.243

\bibitem[Tsuboi et al. 2000]{tsu00} Tsuboi, Y., Imanishi, K., Koyama, K, Grosso, N., \& Montmerle, T.
2000, ApJ, 532, 1089

\bibitem[Weinberger 1998]{wei98} Weinberger, A. 1998, Ph.D. Thesis, Physics, California Institute of
Technology

\bibitem[Wilking, Schwartz, \& Blackwell 1987]{wsb87} Wilking, B.A., Schwartz, R.D., \& Blackwell, J.H.
1987, AJ, 94, 106

\bibitem[Wilking, Lada, \& Young 1989]{wly89} Wilking, B.A., Lada, C.J., \& Young, E.T. 1989,
ApJ, 340, 823

\bibitem[Zhang, Wootten, \& Ho 1997]{zha97} Zhang, Q., Wootten, A., Ho, P.T.P. 1997,
ApJ, 475, 713

\end{thebibliography}
